# Comparative Analysis of Dynamic Data Race Detection Techniques

Danial Entezari

**Abstract**

The consequences of data races can be potentially very problematic [1], and it is important to determine what tools and methods are best at detecting them. The following conditions must be met for a data race to occur: two or more threads in a single process access the same memory location concurrently, at least one of the accesses is for writing, and the threads are not using any exclusive locks to control their accesses to that memory.

This paper reveals the techniques and implementations of the two main methods for dynamic data race detection techniques; the happens-before and lockset methods, and produces an analysis for several tools that employ either (§4, §5) or of both these methods (§7.1) for detecting data races.

This paper also reveals the extent to which dynamic data race detection (also called dynamic analysis) can identify harmful data races, how it can be implemented, and how it compares to other forms of data detection in terms of performance and accuracy.

## 1. Introduction

One major challenge of developing multithreaded programs is the synchronization and concurrency issues that can lead to data races [3]. A data race is caused by poor synchronization implementation in multithread programs, and is a symptom of atomicity violations (Figure 1), ordering violations (Figure 2), unintended sharing (Figure 3), and deadlocks and livelocks (Figure 4) [11].

Data races can be very difficult to detect, especially if the program is large. Often, they manifest only after deployment. A program with data races can be executed repeatedly without revealing any error [15]. Furthermore, the frequency and unpredictability of these bugs will only increase as software adds parallelism to exploit multicore hardware [7].

It is also important to note that data races are not the same as race conditions [28]. A race condition is said to occur when the timing or ordering of events cause the program to produces an error. According to Regehr [28]: "Generally speaking, some kind of external timing or ordering non-determinism is needed to produce a race condition; typical examples are context switches, OS signals, memory operations on a multiprocessor, and hardware interrupts."

To deal with the problem of data races, many tools have been developed for detection using different methods. These methods include: dynamic, post-mortem, and static methods as well as model checking [4]. The dynamic detection method (also called dynamic analysis) detects data races in threaded programs during run time [5]. According to Zhang [2], dynamic data race detectors have usually been based on either a happens-before algorithm or the lockset-based algorithm. Of course, both methods have their own advantages and disadvantages, and a comparison of these techniques is made in this paper §7. Dynamic detection tools may also employ both the happens-before and lockset methods for detecting data races, otherwise known as a hybrid data race detector [6].

To better understand the difference of these two techniques, and to reveal the progress that has been made in the development of tools for detecting and preventing data races, this paper compares happens-before with lockset, illustrates these methods formally, and reviews a selection of tools that fall under the both the lockset method and the happens-before method.

Dynamic data race detection, however, is not the only way to detect data race. This paper also compares static data race detection with dynamic data race detection to make a case for why dynamic data race detection can be a better alternative to static data race detection, to further explore the limitations and capabilities of both techniques, and reveal that both methods can actually be used together - like the Goldilocks and IFRit data race detectors [19, 20, 8].

The motivation for this research was to learn about data races and their detection in greater detail, but also concurrency and synchronization related issues in general. Furthermore, the goal was to discover the techniques available in addressing these issues, and finally, to present these findings to others who may also share an interest in this subject.

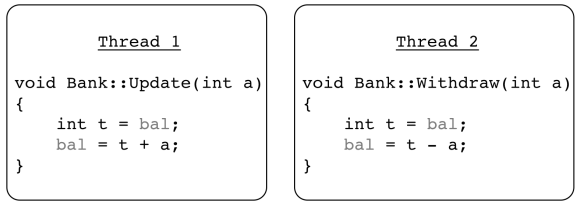

**Figure 1 [11].** Atomicity violation: Thread 2 executes the *Withdraw* method before the *bal* variable is assigned a new value.

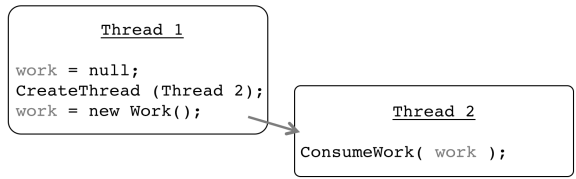

**Figure 2 [11].** Ordering violation: Thread 2 passes the *work* object as a parameter to a function before it is instantiated.

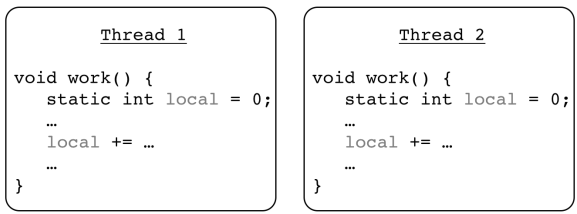

**Figure 3 [11].** Unintended sharing: Thread 1 & 2 concurrently executing the *work* method.

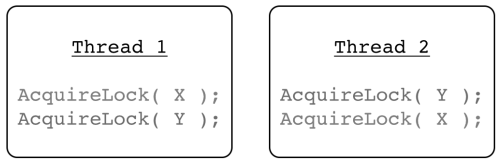

**Figure 4 [11].** Livelock / Deadlock: Threads 1 & 2 have acquired locks the other requires to proceed executing the remainder of their code.

## 2. Methodology

The methodology in this paper involved investigating concurrency and synchronization malpractices that lead to data races and other bugs. The majority of the research literature cited in this paper were focused on new techniques developed to improve on current dynamic data race detection methods. The findings from these research literature led to Lamport's work [18] which happens-before tools are based on, and to the paper on the Eraser detector [3] that first introduced the lockset approach. The happens-before and lockset techniques are discussed in §3. Furthermore, the findings revealed the capabilities of dynamic data race detectors that apply either or both the happens-before and lockset methods, as well as their issues and performance limitations. Happens-before and lockset data race detectors are discussed in sections §4 and §5, respectively.

## 3. Happens-before vs. Lockset

According to Zhang [2] and O'Callahan [15], Dynamic data race detectors have usually been either based on a happens-before algorithm or the lockset algorithm. These authors have described the difference between these techniques as follows: lockset-based tools essentially work based on the principle of locking discipline. They identify a data race if two threads access a shared memory location without holding a common lock. One issue with lockset-based detectors is that they produce too many false positives. Although locks are efficient in providing synchronization, they are not only way to provide safe synchronization and lockset-detectors may not always take this into account [21]. They have the ability to detect data races that could only occur in a particular thread interleaving all by instrumenting a completely different thread interleaving. Compared to happens-before, lockset is more sound but has poor precision. In general, however, it does a good job of detecting most potential data races [21].

Happens-before detectors, on the other hand, identify a data race if two threads access a shared memory location and the accesses are *causally unordered*. Consider two messages m1 and m2. If m1 is sent before m2, then m1 is said to have "happened before" m2, and it is expected that m1 also be received before m2. If this order is maintained, then these event are causally ordered [27]. Therefore, this approach for detecting data races is highly dependent on the actual thread execution ordering that occurs when the instrumented code is run.

Happens-before detectors are more general and can be applied to applications with forms of non-lock-based synchronization, but they are less efficient to implement and are more likely to suffer *false negatives*, i.e., fail to report a data race when there is one, than lockset-based detectors. Happens-before detectors do identify data races that actually exist, however - which means the technique is complete [21].

In general, the warnings that are given by dynamic race detectors operating on the happens-before and lockset algorithms only represent a subset of the actual race conditions in your program; that is, only the executed portion of the program is instrumented - not the entire program.

To formally illustrate the happens-before and lockset algorithm, this paper summarizes the following sections for formal framework (§3.1 and §3.2) from O'Callahan [15]. **Note:** All algorithms, code examples, descriptions, and figures in sections §3.1 and §3.2 are entirely from the O'Callahan [15] paper.

### 3.1 Lockset Algorithm

According to O'Callahan [15], dynamic analysis detects data races by observing a stream of events generated by the instrumentation inserted into a program. O'Callahan [15] presented a framework to allow the happens-before method to be compared to the lockset method. This paper summarizes the framework as follows: A program that is analyzed is treated as an abstract machine that outputs a sequence of events $\langle e \rangle$ to the detector. Each event contains components which is abstracted to the following types:

- $M$: a set of *memory locations*.
- $L$: a set of *locks*.
- $T$: a set of *threads*.
- $G$: a set of *message IDs*
- $A = \{\text{READ}, \text{WRITE}\}$ (the two possible access types for a memory access).

When the program is executed, the following kinds of event are generated. Note: this abstract machine was presented as sequential for simplifying the presentation.

- *Memory access* events of the form MEM($m$, $a$, $t$) where $m \in M$, $a \in A$ and $t \in T$. These indicate that thread $t$ performed an access of type $a$ to location $m$.

- *Lock acquisition* events of the form ACQ($l$, $t$) where $l \in L$ and $t \in T$. These indicate that thread t acquired lock $l$.

- *Lock release* events of the form REL($l$, $t$) where $l \in L$ and $t \in T$. These indicate that thread $t$ released lock $l$ and no longer holds the lock.

- *Thread message send* events of the form SND($g$, $t$) where $t \in T$ and $g \in G$. These indicate that thread $t$ is sending a message $g$ to some receiving thread.

- *Thread message receive*[1] events of the form RCV($g$, $t$) where $t \in T$ and $g \in G$. These indicate that a thread $t$ has received a message $g$ from some sending thread and may now be unblocked if it was blocked before.

The operation of the abstract machine is described by O'Callahan [15] as follows: "At each step, the abstract machine chooses a single thread to run, and executes that thread for some quantum, possibly generating one or more events. Thus events are observed by the detector in a sequence which depends on the thread schedule".

Computation of the set of locks held by a thread at any given time is required before detection. Given an access sequence $\langle e \rangle$, we compute the locks before step $i$ by a thread $t$, $L_i(t)$,

$$\text{as } L_i(t) = \{\, l \mid \exists a. a < i \land e_a = \text{ACQ}(l, t) \\ \land (\nexists r. a < r < i \land e_r = \text{REL}(l, t))\}$$

The "current lockset" for each thread, $L_i(t)$ for each live thread $t$, can be efficiently maintained online (or on-the-fly) as acquisition and release events are received.

A potential race is deemed to have occurred whenever the lockset principle (see §3) is violated. Formally, given an input sequence $\langle e_i \rangle$,

IsPotentialLocksetRace $(i, j) =$
 $e_i = \text{MEM}(m_i, a_i, t_i) \land e_j = \text{MEM}(m_j, a_j, t_j)$
 $\land\, t_i \neq t_j \land m_i = m_j \land (a_i = \text{WRITE} \lor a_j = \text{WRITE})$
 $\land\, L_i(t_i) \cap L_j(t_j) = \varnothing$

O'Callahan [15] presented the code example in Figure 6 [15] to demonstrate data race detection in the lockset method. The statement *childThread.interrupt()* generates a memory access with location *main.childThread*, type READ, thread MAIN, and lockset *{main}*. The statement *main.childThread* generates a memory access on *main.childThread* with type WRITE, thread CHILD and lockset $\varnothing$. Therefore IsPotentialLocksetRace will be true for these two events.

### 3.2 Happens-before Algorithm

According to O'Callahan [15], The happens-before relation was defined by Lamport as a partial order on events occurring in a distributed system. The basic idea of the happens-before relation is that a pair of events ($e_i$, $e_j$) are related if communication between processes allows information to be transmitted from $e_i$ to $e_j$. In Figure 5 [15], event E1 happens-before event E2, but E2 and E3 are unrelated. Given an event sequence $\langle e \rangle$, the happens-before $\rightarrow$ relation is the smallest relation satisfying the following conditions:

---

[1] Thread message events are only observed by the happens-before detector [15]

- If $e_i$ and $e_j$ are events in the same thread, and $e_i$ comes before $e_j$, then $i \rightarrow j$. Formally,

$$\text{Thread}(e_i) = \text{Thread}(e_j) \wedge i < j \Rightarrow i \rightarrow j$$

- If $e_i$ is the sending of message $g$ and $e_j$ is the reception of $g$, then $i \rightarrow j$.

$$e_i = \text{SND}(g, t_1) \wedge e_j = \text{RCV}(g, t_2) \Rightarrow i \rightarrow j$$

- Happens-before is transitively closed.

$$i \rightarrow j \wedge j \rightarrow k \Rightarrow i \rightarrow k$$

Consider Figure 6 [15]. The only inter-thread happens-before relationships that will be produced are that *globalFlag = 1* happens-before *if (globalFlag)* and the following statements, and that *childThread.start()* happens-before *ChildThread.run()* and the following statements. Therefore a happens-before detector will never report a race on globalFlag, but it will report a race between *main.childThread = null* and *if (childThread != null)*.

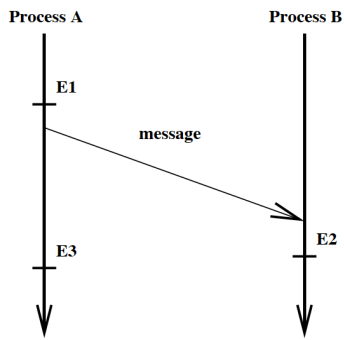

**Figure 5 [15].**

## 4. Techniques in Happens-before

Formulated by Leslie Lamport [18], the happens-before relation, which these techniques are based on, will determine from the results of events the actual order in which they occurred. In other words, if event a occurred before b, then the results of these events should reveal the order of their occurrence.

### 4.1 FastTrack:

FastTrack is a happens before data race detector that employs vector clocks to track threads and locks. Vector clocks are precise but slow. FastTrack optimizes VC-based detection via *adaptive representation*; that is, replacing full vector-clocks with *epochs*, and using full vector-clocks when needed.

A vector clock is an integer vector that is used to keep track of segments of different threads. These segments are then compared to one another to determine if there are data races [13]. FastTrack itself is based on the happens-before Djit+ algorithm [22]. VC-based race detectors are considered to be precise race detectors because they never produce false alarms and they can track the happens-before relationship. VC-based detectors, however, incur significant performance overhead because they record information about each thread in a system.

This is where FastTrack improves on VC-based detection. The authors of FastTrack argue that it is not necessary for each and every thread to be instrumented - which is what VC-based detectors generally do. According to O-K. Ha, & Y-K. Jun [10], the Djit+ algorithm, that the FastTrack is based on, exploits the idea that all writes to a variable are totally ordered by the happens-before relation and so it records only the most recent writes.

```
// MAIN THREAD
class Main {
  int globalFlag;
  ChildThread childThread;
  void execute() {
    globalFlag = 1;
    childThread = new ChildThread(this);
    childThread.start();
    ...
    synchronized (this) {
      if (childThread != null) {
L:      childThread.interrupt();
      }
    }
  }
}

// CHILD THREAD
class ChildThread extends Thread {
  Main main;

  ChildThread(Main main) { this.main = main; }
  void run() {
    if (main.globalFlag == 1) ...;
    ...
    main.childThread = null;
  }
}
```

**Figure 6 [15].**

The FastTrack algorithm uses what it calls *adaptive representation* techniques to provide fast paths in constant time, without losing the precision and correctness associated with vector-clock race detection. FastTrack expands on the Djit+ approach of recording the most recent writes to a variable by a thread in that it records only the very last write - specifically, the clock and thread identifier of that write. This tuple pair of a clock and a thread identifier is referred to as an *epoch*. Fast-Track, then, replaces heavyweight vector-clocks with epochs, and this decreases the runtime and memory overhead of almost all VC operations from $O(n)$ to $O(1)$ in the detection of data races, where $n$ designates the maximum number of simultaneously active threads during an execution. Figure 7 describes the formal definition of an epoch.

We refer to a pair of a clock c and a thread t as an epoch, denoted c@t.

An epoch $c@t$ *happens before* a vector clock $V$ ($c@t \preceq V$) if and only if the clock of the epoch is less than or equal to the corresponding clock in the vector.

$$c@t \preceq V \quad \text{iff} \quad c \leq V(t)$$

**Figure 7 [9].**

The adaptive representation referred to by the authors is the way that FastTrack "switches from epochs to vector clocks where necessary (for example, when data becomes read-shared), and from vector clocks back to lightweight epochs where possible (for example, when read-shared data is subsequently updated)". This, of course, is done to retain the precision of the algorithm in detecting data races.

According to the authors [9], synchronization operations (lock acquires and releases, forks, joins, waits, notifies, etc) account for a very small fraction of the operations that must be monitored by a race detector. Reads and writes to object fields and arrays, on the other hand, account for over 96% of monitored operations. This information is based on empirical data collected from a variety of Java programs.

### 4.2 Portend

It is important for a data race detection tool to classify data races according to their effects to help programmers to more efficiently address harmful data races in their programs.

Portend is a tool based on a happens-before algorithm that not only detects data races dynamically but also classifies them based on the consequences they could potentially incur. While it's true that data races are a sign of poor or faulty synchronisation practices, and they could potentially have serious consequences, the authors of this paper argue [1] that it is still not necessary to address every single data race. Firstly, synchronizing all data races would introduce unacceptable levels of performance overhead, and secondly, most data race (76% - 90%) are simply harmless. Instead of overwhelming developers and resources with data race detection, Portend's goal is to identify and report data races that are actually harmful. For this, Portend introduces a data race classification theme. According to the paper with regard to the classification scheme, a simple "harmless" vs "harmful" is not helpful. Therefore, a scheme that more precisely can describe what is identified is required. This classification scheme classifies the true races into four categories: "spec violated", "output differs", "k-witness harmless", and "single ordering". This taxonomy is illustrated in Figures 10 and 11.

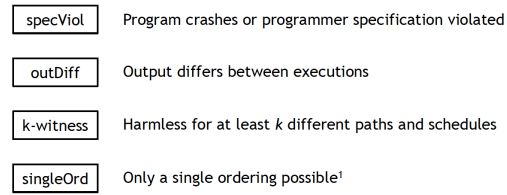

| specViol | Program crashes or programmer specification violated |
| outDiff | Output differs between executions |
| k-witness | Harmless for at least *k* different paths and schedules |
| singleOrd | Only a single ordering possible[1] |

[1] Arguably not a data race

**Figure 10 [1].**

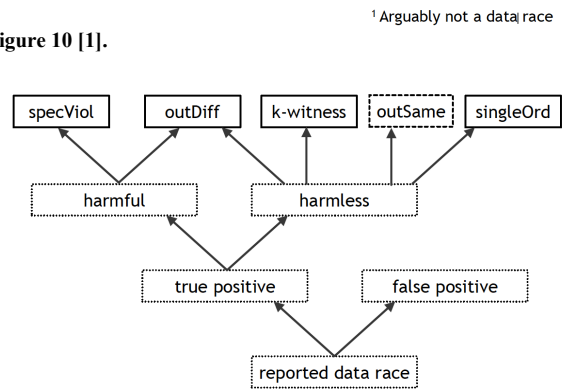

**Figure 11 [1].**

When Portend identifies a data race as harmful, it produces re-playable traces (i.e., inputs and thread schedule) that will demonstrate to the programmer what the harmful effects of the data race will be. This will make it easier for the programmer to resolve the issue with the data race. Portend's reports also include stack traces of the threads causing data race along with the size of the accessed memory field, and stack races for segmentation faults. Portend operates on binaries, not on source code since it is a dynamic data race detector.

To detect data races in a program, the program must be fed to the Portend race detector for analysis and reporting. The data race detection is a multi-step process that begins with the single-path/single-schedule analysis as its first step, followed by the second step, multi-path analysis, and symbolic output comparison augmented with multi-schedule analysis.
The goal of the first step is to identify cases in which the alternate schedule of a race cannot be pursued, and to make a first classification attempt based on a single alternate execution. The next step, multi-path analysis, will explore variations of the single paths found in the previous step (i.e, the primary and the alternate) in order to expose Portend to a wider range of execution alternatives.

## 5. Techniques in Lockset

The lockset algorithm was first introduced and implemented by the Eraser data race detector, the first lockset technique this paper reviews in this section. The authors of this paper argued that the generality of Lamport's happens-before was costly, and Eraser's approach - the lockset approach - would be simpler, more efficient, and more thorough at catching races than happens-before data race detectors.

### 5.1 Eraser

Eraser uses binary rewriting techniques to monitor every shared memory reference and verify that consistent locking behaviour is observed. It works by monitoring data accessed at a very low level and observing patterns [12].

Eraser monitors every shared memory references to verify that locks are used when accessing and/or writing to shared data to prevent data races. Eraser relies on the execution history to infer the protection relation of locks; that is, which lock is protecting what variable. Eraser employs 2 algorithms: The first is a simple algorithm, and the other is an improved algorithm. Eraser is based on the lockset algorithm as described §3.2.

The Eraser paper describes this algorithm as "too-strict" because there are programming practices that violate the locking discipline of the Eraser algorithm. These practices, however, never produce race conditions. These very common programming practices are described as follows:

- Initialization: Shared variables are frequently initialized without holding a lock.

- Read-Shared Data: Some shared variables are written during initialization only and are read-only thereafter. These can be safely accessed without locks.

- Read-Write Locks: Read-write locks allow multiple readers to access a shared variable, but allow only a single writer to do so.

Unlike lockset algorithms, Happens before algorithms handle many styles of synchronization, but this generality comes at a cost. Eraser is aimed specifically at the lock-based synchronization used in modern multithreaded programs. Eraser simply checks that all shared-memory accesses follow a consistent locking discipline. A locking discipline is a programming policy that ensures the absence of data races. For example, a simple locking discipline is to require that every variable shared between threads is protected by a mutual exclusion lock. This is as described in the happens-before vs. lockset overview §3. The authors argued that the lockset approach the Eraser had introduced and implemented would be simpler, more efficient and more thorough at catching data races than the happens-before approach.

### 5.2 PECAN

Data race detection that produces too many false positives is non-trivial and not helpful for programmers. It is important for data race detection tool to help the programmer to understand the cause and harmful effects of bugs.

The authors of this paper [16] introduce a technique and a prototype tool that detects bugs called PECAN (Persuasive Prediction of Concurrency Access Anomalies). The aim of PECAN is to help programmers to find harmful bugs, including harmful data races, and also inform the programmer as to what caused the bug to occur.

According to the authors: "PECAN uses predictive trace analysis or PTA. Generally speaking, a PTA technique records a trace of execution events, statically (often exhaustively) generates other permutations of these events under certain scheduling constraints, and exposes concurrency bugs unseen in the recorded execution". Furthermore, as stated by the authors: "PTA is a powerful technique as, compared to dynamic analysis, it is capable of exposing bugs in unexercised executions and, compared to static analysis, it incurs much fewer false positives for the fact that its static analysis phase uses the concrete execution history" [16].

Concurrent programs are plagued by a class of bugs called access anomalies (AAs), characterized by criteria such as data race, atomicity violation, and atomic-set serializability violations.

Just as with Portend §4.3, the authors of this paper argue that a bug detection technique is more useful if it is informative, or as the authors put it - more "persuasive". This new criterion emphasizes that a bug detection technique should not only localize the bug in the source code but also, and more importantly, help programmers in fully understanding how the bug has occurred, to provide good fixes.

The primary goal of PECAN, then, is providing data race reports that would persuade the programmer to resolve the issue. The approach of PECAN is two fold: First, the report should not include any false positives as too many false positive reports will, perhaps, overwhelm the programmer. As a result, the programmer may forget or choose to ignore the report. This is the first persuasive technique. The paper adds that "Since it is non-trivial to manually verify the false alarms in large sophisticated concurrent systems, the perceived usefulness of the technique quickly deteriorates with even a small number of false positives". Second, the report should describe to the programmer how the detected bug or violation can occur. This is done by providing with each report a bug-hatching clip. A bug hatching clip instructs the program to deterministically execute the bug hatching process in steps. In other words, the programmer can replay the program and produce the same bug again and again.

## 6. Hybrid Techniques

According to O'Callahan [15], Implementing one algorithm may not suffice. Happens before detection can be costly in that vector clocks for every shared memory location and every locks must be maintained, and fewer bug will be found. Lockset based detection does not consider (generally) other forms of safe synchronisation practices and produces too many false positives. To benefit from the precision of happens-before race detection and the low overhead performance of lockset race detection, ThreadSanitizer is an tool that implements both methods.

### 6.1 ThreadSanitizer

ThreadSanitizer, also known as TSAN, is tool that is uses both the lockset and happens-before algorithms to instrument source code and identify data races in a program, and to improve precision in detecting data races and reduce false positives.

TSAN's algorithm has several modes of operation. In its conservative mode, it reports fewer false positive but is not as precise. In its very aggressive mode, it is more precise but reports more false positives.

There are two versions of TSAN; TSAN v1 and TSAN v2. TSAN V1 is implemented in Valgrind, a heavyweight binary instrumentation framework. TSAN v2 is a compiler-based instrumentation pass in LLVM [25].

In compile instrumentation, TSAN intercepts all reads and writes. This is done by inserting function calls wherever data race issues may potentially occur (Figure 12).

According to the authors, dynamic data race detectors must understand the synchronization mechanisms used by the program is it detecting. TSAN uses *dynamic annotations*, a "kind of race detection API" to understand the synchronization used by the program it is testing. Furthermore, each dynamic annotation is a C macro definition that is expanded into code that is intercepted and interpreted by TSAN.

```
void foo(int *p) {
  *p = 42;
}

         ⬇

void foo(int *p) {
  __tsan_func_entry(__builtin_return_address(0));
  __tsan_write4(p);
  *p = 42;
  __tsan_func_exit()
}
```

**Figure 12 [26].**

## 7. Dynamic vs. Static Analysis

While the focus of this paper is dynamic analysis techniques for detecting data race, this paper, however, also explores another alternative to this approach - static analysis. This papers compares static analysis to dynamic analysis for two reasons: the first reason is to further define the abilities and limitations of both techniques (especially dynamic analysis). Second, both techniques when used together can further improve the process of detecting data race, and this is revealed in the Goldilocks (§7.1) and IFRit (§7.2) reviews.

In general, there are two methods for detecting data races: dynamic analysis and static analysis. Dynamic analysis will detect data races by executing a program and analyzing the results. A dynamic test, however, will only find defects in the part of the code that is actually executed [29]. Static analysis will detect races by analyzing the source codes of programs [17]. It explores a simplified version of the concrete program, examining the entire behavior. It will inspect program code for all possible run-time behaviors and seek out coding flaws, back doors, and potentially malicious code [29].

The advantage of dynamic analysis over static analysis is that dynamic analysis does not produce as many false positives, and annotations are not required by the programmer to for helping the detector in detecting data races. On the other hand, static analysis does not incur any performance overhead and can reason over all interleaving [11].

Static analysis and dynamic analysis need not be mutually exclusive, however. In fact, it may be more effective to combine both methods [15, 19, 20, 6]. This paper reviews two dynamic detection tools that also utilize static analysis to improve data race detection results.

### 7.1 Goldilocks

In the happens-before vs. lockset comparison (§2), this paper established that while lockset-based technique are more efficient, they are less precise compared to their counterparts that implement the happens-before technique. To address this, Goldilocks, a lockset based tool implemented for Java, accomplishes two things: firstly, it further improves the lockset efficiency, and secondly, it improves precision of lockset detection. Besides applying dynamic analysis to find data races, Goldilocks also applies static analysis to reduce the runtime overhead of data race detection.

According to S. Adve [24], Goldilocks treats data races as language level exceptions, just like null pointer dereferences. The authors of Goldilocks [19] state that data races complicate the semantics of a program. If the semantics are complicated, the program may become sequentially inconsistent, and this is where data races thrive. The approach Goldilocks takes, then, is executions are either sequentially consistent or there will be a thrown exception. This approach, as described by S. Adve [24], resolves the most difficult part of the memory models mess.

To improve the precision of lockset detection, Goldilocks combines the precision of vector clocks with the computational efficiency of locksets to captures the happens-before relation.

### 7.2 IFRit

IFRit uses static analysis to identify interference-free regions that reduce redundant instrumentations at compile time [23]. As a result, there is significant reduction in overhead during analysis.

IFRit improves on the happens-before approach in that it does not report false positives - which is what detectors based on the happens-before approach generally do [2]. Also, it does not instrument every memory access, and it does not track a full happens-before relation. As a result, it greatly reduces overhead associated with happens-before detection techniques without completely losing precision (though, it may still miss data races).

According to the authors [8], one way to resolve the issue of happens-before detectors producing too many false positives is sampling; that is, removing instrumentation from some memory accesses. IFRit expands on this idea of not requiring complete instrumentation for each and every memory access. To demonstrate this idea, the paper describes the following example: Consider the code in Figure 8 containing a for loop after the 'lock' statement and before the 'unlock' statement. This region contains synchronization (as an assumption) and so is considered to be *synchronization-free*. The authors state that it is not necessary to instrument each access to $x$ in this example. Instead, to detect data races on the variable $x$, it is sufficient to instrument the region between the 'lock' and 'unlock' statements as 'writing to $x$', which dynamically requires instrumentation only before and after the loop.

```
lock(m);
for(int i = 0; i < 1000; i++) {
  ...
  *x = ... ;
  ...
}
unlock(m);
```

Figure 8 [9].

This dynamic data-race detector, is based on a run-time abstraction called an interference-free region (IFR). An interference-free region is defined to be a generalization of a synchronization-free region; it is a specific region around a memory access that can include synchronization release operations before the access and synchronization acquire operations after the access (Figure 9). If another thread has a conflicting memory access that is unsynchronized with respect to the interference-free region, then the accesses must form a data race [14]. Interference-free regions are so called because while the thread is executing the interference-free region, no other thread can write to the shared variable accessed by the IFR's access without inducing a data race.

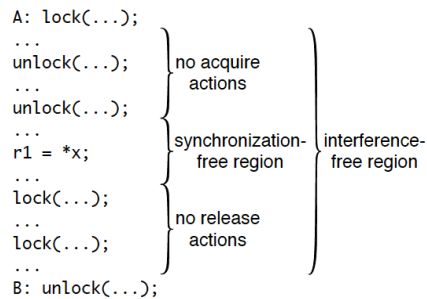

Figure 9 [9].

The way IFRit determines that a data race will occur is this: For example. there are two accesses to the same variable (the variable is in different threads). If the

IFRs of these accesses overlap; that is, part of their executions happens simultaneously, and at least one of these IFRs is writing to the variable, then these accesses are going to cause a data race.

## 8. Conclusion

This paper revealed that there are generally two methods to dynamically detect data races (also called dynamic analysis); namely, the happens-before and the lockset techniques, and dynamic data race detection can employ either or both algorithms. These algorithms were presented formally and examples of how both methods identify data races were provided[2]. This was followed by a general review of happens-before detection techniques [9, 1] and Lockset based detection techniques [3, 16]. This was done to gain better familiarity with the advantages and disadvantages of both methods. As multithreaded systems continue to evolve, it becomes very crucial for developers to engage in proper multithreaded programming practices, but more importantly, have available to them efficient and precise tools that help them identify bugs. This paper provided a summary of data race detection techniques with the goal of providing its readers with a broad overview on the subject of dynamic data race detection, and perhaps, helping them to find the right tools and techniques to address concurrency and synchronization related issues in their programs. The paper also discussed alternative ways to detecting data races, like static data race detection (also called static analysis), and reviewed two examples of tools that employ both dynamic and static data race detection (§7.1 & §7.2).

This research revealed the following about data race detectors that were discussed: Vector-clocks, while precise, can be very slow unless there is a substitute for full generality for instrumentation [9]; Instrumentation to every memory access is not always necessary and accuracy will not be sacrificed as a result [14]; It is important for a data race detection tool to classify bugs (and especially data races) to make it more feasible for programmers to remove harmful data races [1]; binary rewriting, an approach in lockset-based detection, can be used at the very low level to dynamically detect data races and this can be more efficient (and still as precise) as happens-before detection [3]; Allowing programmers to deterministically replay the bugs in their program will make it easier and more "persuasive" for programmers to resolve concurrency and synchronization issues [16]; happens-before and lockset methods can actually be used together to dynamically detect data races [6]; and finally, dynamic analysis is not the only way to detect data races - static analysis and dynamic analysis can also be combined for effective data race detection [19, 20].

## 9. Related Work

The findings of this paper are generally borrowed from investigations on data race detection techniques by the papers referenced in this research. With respect to related work on findings on the happens-before and lockset techniques, the paper by O'Callahan [15] presents algorithms, performance results, and code examples for each case, in addition to introducing a hybrid algorithm that improves on the state of the art in accuracy, in usability, and in overhead.

The paper by C. Sadowski [25] investigates how two program analysis tools are used by developers at Google: ThreadSafety, an annotation-based static data race analysis, and TSAN, a dynamic data race detector. The data was collected by interviewing seven veteran industry developers at Google, and provides unique insight into how four different teams use tooling in different ways to help with multithreaded programming.

This paper itself has not provided performance results, however. The works cited in this paper generally provide performance results and in-depth analysis of the tools they have presented. For related work that have produced comparisons results in performance and other criteria, the IFRit paper [8], provides sampling and overhead comparison results for FastTrack [9], ThreadSanitizer [6], and IFRit itself; the FastTrack paper [9] provides performance results for instrumentation time, vector-clock allocation, and warning for dynamic data race detectors, including Eraser and Goldilocks.

---

[2] Formal presentation of happens-before and lockset algorithms, including code examples and figures are entirely from O'Callahan [15].